# International evidence on business cycle magnitude dependence

Corrado Di Guilmi,[a] Edoardo Gaffeo,[b, *]
Mauro Gallegati,[a] Antonio Palestrini[c]

[a] *Department of Economics, Università Politecnica delle Marche, Piaz.le Martelli 8, I-62100 Ancona, Italy*
[b] *Department of Economics, University of Trento, Via Inama 5, I-38100 Trento, Italy*
[c] *Department of Law in Society and History, University of Teramo, Via Crucioli 120, I-64100 Teramo, Italy*

December 18th, 2003

**Abstract**

Are expansions and recessions more likely to end as their magnitude increases? In this paper we apply parametric hazard models to investigate this issue in a sample of 16 countries from 1881 to 2000. For the total sample we find evidence of positive magnitude dependence for recessions, while for expansions we are not able to reject the null of magnitude independence. This last result is likely due to a structural change in the mechanism guiding expansions before and after the second World War. In particular, upturns show negative magnitude dependence in the post-World War II sub-sample, meaning that in this period expansions become less likely to end as their magnitude increases.

• Corresponding author.
Tel. +39 (0)461882220; Fax +39 (0)461882222; E-mail: edoardo.gaffeo@economia.unitn.it.

# 1. Introduction

Great effort has been put forth so far by the profession to investigate whether the business cycle exhibits duration dependence. In fact, the received wisdom in mainstream macroeconomic theory is that business fluctuations are driven by recurring identically independently distributed (*iid*) random shocks, so that cycles' duration should be independent of length. The empirical work on duration dependence has been conducted by means of both nonparametric (McCulloch, 1975; Diebold and Rudebusch, 1990) as well as parametric techniques (Sichel, 1991; Diebold *et al*., 1993; Zuehlke, 2003), the latter being generally favoured because of modelling convenience and higher statistical power. While the evidence as a whole is far from conclusive, a relative consensus has been recently established in favour of positive duration dependence, at least for pre-war expansions and post-war contractions. A tentative explanation for this result has been advanced by Sichel (1991). Suppose that policymakers are interested in lengthening expansions. Hence positive duration dependence of contractions and null duration dependence of expansions might emerge simply because policymakers are urged to act countercyclically as downturns length, while macroeconomic policy mistakes could be as likely to end short as well as long expansions.

This type of reasoning resembles the so-called stabilization debate, that is whether macroeconomic policy effectiveness in decreasing the volatility of business cycle fluctuations around trend has improved after the second World War (WWII). See e.g. Romer (1999) for a paper arguing in the affirmative, and Stock and Watson (2003) for a partially dissenting view. Regardless of the position one is prone to take in this debate, it should be recognized that policymakers are plausibly not interested in the length of a business cycle phase (to end it if a recession, and to lengthen it if an expansion) per se. A recession could be disturbing not only if it lengthens a lot, but also if it is short *and* particularly severe. By the same token, an expansion could force antinflationary (i.e., countercyclical) policies if it is gaining excessive momentum, regardless of its duration. In other terms, macroeconomic policy could well be targeted at controlling the magnitude of business cycle phases, rather than their duration alone.

In line with this assumption, in this paper we aim to extend the empirical literature on dependence in business cycles by posing a related but different question: are expansions or contractions in economic activity more likely to end as they become bigger? We apply parametric techniques for a pool of 16 countries sampled from 1881 to 2000. For the whole sample, we find evidence of positive magnitude dependence for contractions, while expansions do not exhibit any significant dependence, a result which seems to confirms the behavioural hypothesis proposed by Sichel. This result has been further scrutinized by conducting parameter estimates for the pre-WWII and the post-WWII sub-sample separately. Interestingly, while for contractions the same result holds regardless of the sample used, for expansions we find positive magnitude dependence for the pre-WWII period, but negative magnitude dependence for the post-WWII era. In other terms, if policymakers' increased activism in the years following WWII has to be praised for the evidence at hand, our findings suggest that macroeconomic policies have been applied more sensibly in magnifying the magnitude of expansive phases.

The paper is organized as follows. A measure of magnitude of business cycle phases and parametric methodologies to investigate their dependence are discussed in section 2, while empirical results are presented in section 3. Section 4 concludes.

## 2. Methodology

The concept of business cycle fluctuations we adopt here is that of growth cycles, that is expansions and contractions expressed in terms of deviations from an estimated GDP trend or potential (Zarnowitz, 1992). A useful method to summarize information on either time (i.e., duration) and size (i.e. output gap) of any single episode consists in calculating its *steepness*,[1] expressed as the ratio of the amplitude *y* (i.e. cumulative percentage points of peak-to-trough and trough-to-peak output gap for recessions and expansions, respectively) to the amplitude *t* (in time periods), $x = \frac{y}{t} > 0$.[2] In what follows, we will use this measure to approximate the magnitude of a business cycle phase.

The following step consists in deriving the conditional probability that phase *i* will end at magnitude $x_i$, given that a magnitude $x_i$ has been obtained. Our investigation is based on a Weibull parametric hazard model (Lancaster, 1979):

$$h_W(x) = \alpha \beta x^{\beta-1} \qquad (1)$$

and on its associated survivor function $S^W(x) = \exp(-\alpha x^\beta)$, with $\alpha = \eta^{-\beta}$, $\eta > 0$ being the *scale* parameter, while the *shape* parameter $\beta > 0$ measures the elasticity of magnitude dependence. If $\beta$ is higher (lower) than one, then the conditional probability that a phase ends increases (decreases) linearly as its magnitude increases. Finally, when $\beta$ is equal to 1, the hazard

---

[1] The concept of steepness we use has a geometrical meaning and it is therefore different from the one in Sichel (1993), where this same term has been used to define a property of asymmetric business fluctuations.

[2] This measure corresponds to the slope of the hypotenuse of the *triangle approximation to the cumulative movement* during a business cycle phase as discussed in Harding and Pagan (2002).

function corresponds to that of an exponential, or memory-less, distribution.

It is well known that in model (1) unobserved heterogeneity across observations biases the estimate of the elasticity parameter β downward (Lancaster, 1979). In particular, if estimates point towards negative magnitude dependence it could be practically impossible to identify whether such a result owes to true negative dependence or to heterogeneity bias. McDonald and Butler (1987) explain how to use mixture distributions to handle heterogeneity, showing that if the location parameter is inverse generalized gamma distributed, the distribution for observed data will be Burr type IIX.

As discussed in section 3 below, our empirical exercise is based on a sample of pooled international data, so that heterogeneity is likely to seriously affect our estimates. Hence, in addition to the standard 2-parameter Weibull model (*W*) we check the robustness of our results by recurring to the hazard function for the generalized 3-parameter Weibull model proposed by Mudholkar *et al*. (1996) (*MSK*), which contains the Burr type XII distribution as a special case:

$$h_{MSK}(x_i) = \alpha\beta x^{\beta-1} [S_{MSK}(x_i)]^{-\gamma} \qquad (2)$$

where $S_{MSK} = [1 - \gamma\alpha x^\beta]^{\gamma^{-1}}$ is the corresponding survivor function, the *location* parameter γ is real, and the sample space for *x* is (0, ∞) for γ < 0 and $\left(0, (\alpha\gamma)^{-(\beta-1)^{-1}}\right)$ for γ > 0. Besides correcting for unobserved heterogeneity, the additional parameter γ allows the hazard function to be nonlinearly monotonic increasing (β > 1, γ > 0), nonlinearly monotonic decreasing (β < 1, γ < 0), bathtub shaped (β < 1, γ > 0), unimodal (β > 1, γ < 0) or constant (β = 1, γ = 0). Finally, when β > 0 and γ ≤ 0 the generalized Weibull corresponds to the Burr type XII distribution.

For both models parameters' estimation has been conducted by means of Maximum Likelihood. The log-likelihood function for a series of expansions (contractions) with observed magnitude $x_i$ is:

$$\ln L_j(\bullet) = \sum_{i=1}^{N} \left\{ \Lambda_i \ln[h_j(x_i)] + \ln[S_j(x_i)] \right\} \qquad (3)$$

with $j = W$, $MSK$, and where $\Lambda_i$ is a binary variable that controls for the censoring of incomplete phases. Given that we are operating with nested models, a significantly positive (negative) estimate of $\gamma$ is evidence, besides of positive or bathtub shaped (negative or unimodal) magnitude dependence, of a statistical superiority of the $MSK$ model relative to the $W$ model (Zuehlke, 2003). Furthermore, the sign of the estimated $\gamma$ allows us to control for heterogeneity in the data: the magnitude elasticity obtained with the $W$ model is likely to be biased downward as soon as $\gamma$ in the $MSK$ model is significantly negative.

## 3. Empirical results

The data we use are annual GDP indexes for 16 countries[3] spanning from 1881 through 2000 (IMF, 2002). The time series do not contain data for the periods corresponding to the two WWs, i.e. 1914-1919 and 1939-1948. For each country, the GDP potential has been calculated by means of the Hodrick-Prescott low-pass filter (Hodrick and Prescott, 1997), with the smoothing parameter $\lambda$ equal to 100. Finally, in order to obtain enough observations to attain reliable estimates, we built samples for expansions

---

[3] The 16 countries are Australia, Canada, Denmark, Finland, France, Germany, Italy, Japan, Netherlands, Norway, Portugal, Spain, Sweden, Switzerland, United Kingdom and United States.

(276 observations) and contractions (284 observations) by pooling data for individual countries[4].

Given that both parameterizations (1) and (2) yields hazard functions belonging to the Weibull family, as a preliminary check of model adequacy we study the magnitude empirical distribution of pooled expansions and contractions by means of Weibull probability plots. An advantage of such a plotting technique is that it allows to gain insights on the appropriateness of a distribution model by visual inspection: if the data come from a Weibull distribution the plot will be linear, while other distribution types would introduce curvature in the plot. Figure 1 shows that our modelling strategy finds a convincing support for contractions (Panel *b*), with observations distributed along the reference line but for few data points in the left tails, and a reasonable support for expansions (Panel *a*), with the Weibull model yielding a good fit to the empirical distribution for the central portion, but a relatively poor fit for both tails.

In Table 1 we report the Maximum Likelihood parameter estimates, along with their asymptotic standard errors, obtained with the *W* model (1) and the *MSK* model (2) for the full sample, the pre-WWII sub-sample and the post-WWII sub-sample, respectively. Expansions and contractions are treated separately.

---

[4] This allows us to employ a number of observations one order of magnitude higher that the ones usually employed in the duration dependence literature.

Table 1. *Tests of magnitude dependence in pre-WWII, post-WWII and total sample expansions and contractions for a pool of 16 countries (1881-2000). Parameter estimates have been obtained by ML for the W model (1) and for the MSK model (2). Standard errors in round brackets, p-values in square brackets.*

|  | Expansions | | | Contractions | | |
|---|---|---|---|---|---|---|
|  | Total sample | Pre-WWII sub-sample | Post-WWII sub-sample | Total sample | Pre-WWII sub-sample | Post-WWII sub-sample |
| 1) *W* model | | | | | | |
| $\eta$ | 0.0387† | 0.0517† | 0.0259† | 0.0401† | 0.0533† | 0.0265† |
|  | (0.0024) | (0.0028) | (0.0027) | (0.0018) | (0.0024) | (0.0017) |
| $\beta$ | 1.0143 | 1.4440** | 0.8643** | 1.3832** | 1.6792** | 1.5173** |
|  | (0.0384) | (0.0753) | (0.0444) | (0.0604) | (0.0912) | (0.1007) |
| 2) *MSK* model | | | | | | |
| $\eta$ | 0.0358† | 0.0496† | 0.0242† | 0.0389† | 0.0505† | 0.0257† |
|  | (0.0023) | (0.0028) | (0.0026) | (0.0018) | (0.0024) | (0.0017) |
| $\beta$ | 0.9635 | 1.3853** | 0.8402** | 1.3314** | 1.5808** | 1.4678** |
|  | (0.0375) | (0.0730) | (0.0437) | (0.0586) | (0.0869) | (0.0979) |
| $\gamma$ | 0.0018† | 0.0016 | 0.0011† | 0.0009 | 0.0022 | 0.0006 |
|  | [0.0000] | [0.4091] | [0.0004] | [0.2259] | [0.4144] | [0.6995] |

** Significantly different from unity at the 5% level using a one-tailed test.
† Significantly different from zero at the 5% level using a one-tailed test.

The evidence from the *W* model shows that, for the total sample, positive magnitude dependence exists for contractions, while for expansions we are not able to reject the null of magnitude independence at standard significance levels with a one-tailed test. This last result occurs because of a structural change over the period studied. Pre-WWII expansions exhibit magnitude dependence ($\beta$ = 1.444), while for the post-WWII sample the dependence elasticity is lower than one ($\beta$ = 0.8643), meaning that in this case the probability expansions end decreases with their magnitude. In turn, contractions exhibit a substantially similar degree of positive magnitude dependence either in the pre-WWII ($\beta$ = 1.6792) and in the post-WWII samples ($\beta$ = 1.5173).

Estimates for the *MSK* model seems to confirm the robustness of our findings. The additional location parameter $\gamma$ turns out to be positive but very small in all cases. In this case, an assessment of the statistical superiority of the *MSK* model relative to the *W* model cannot be based on standard tests build on asymptotic standard deviations because of their low power. Instead, a better strategy consists in using a *minimum statistic* test under the null $\gamma = 0$, which returns the probability to observe a minimum above the ML estimate of $\gamma$.

Contractions still show positive magnitude dependence both in the pre-WWI and in the post-WWI era. As regards expansions, positive magnitude dependence is detected in the pre-WWII period, while for the post-WWII period the parameters estimates suggest a bathtub shaped hazard function. In fact, over the range of variation of our data the degree of non-linearity introduced by the *MSK* model is negligible for any practical purpose, as one can realise by visually inspecting the *MSK* hazard plots for expansions and contractions shown in Figure 2.

## 4. Conclusion

This paper presents evidence on magnitude dependence of business fluctuations in a pooled sample of 16 industrialized countries from 1881 through 2000. Drawing from the literature on duration dependence (Sichel, 1993; Zuehlke, 2003), we employ a Weibull parametric model to detect positive magnitude dependence for contractions both in the pre-WWI and in the post-WWII periods, while expansions exhibit positive magnitude dependence in the pre-WWII period, and negative magnitude dependence for the post-WWII sub-sample. These findings do not seem to be affected by heterogeneity in the data, nor are they qualitatively changed as we perform estimates with the generalized model proposed by Mudholkar *et al*. (1996).

There many possible explanations for the evidence at hand. Among them, the most appealing for us has to do with the so-called stabilization debate, that is whether the coming out of automatic stabilizers and the increased ability in conducting monetary policy after WWII has significantly contributed to decrease volatility in aggregate economic activity. Our starting point is that stabilization macroeconomic policy is generally aimed at affecting either the duration and the deepness of business cycle fluctuations: policymakers are better off if sustainable (i.e., without significant inflationary pressures) expansions lengthen a lot, and mild recessions are short. The measure we use for the magnitude of a business cycle phase, i.e. steepness, is designed precisely to capture both aspects. From this perspective, the structural shift we find for expansions before and after WWII could be interpreted as an indirect evidence that macroeconomic policy has became more effective from the 1950s on.

## Acknowledgements
Many thanks to James Morsink for the data, and to Jack Lucchetti for helpful conversations. The usual disclaimers apply.

Figure 1. *Weibull probability plots for expansions and contractions, full sample.*

*a) Expansions*

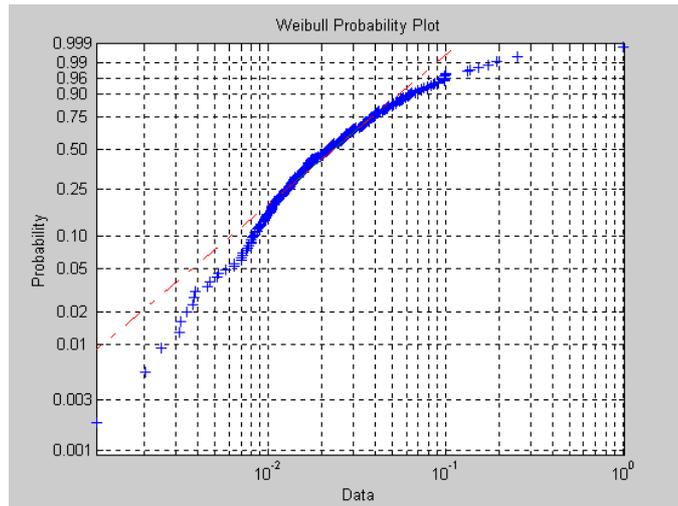

*b) Contractions*

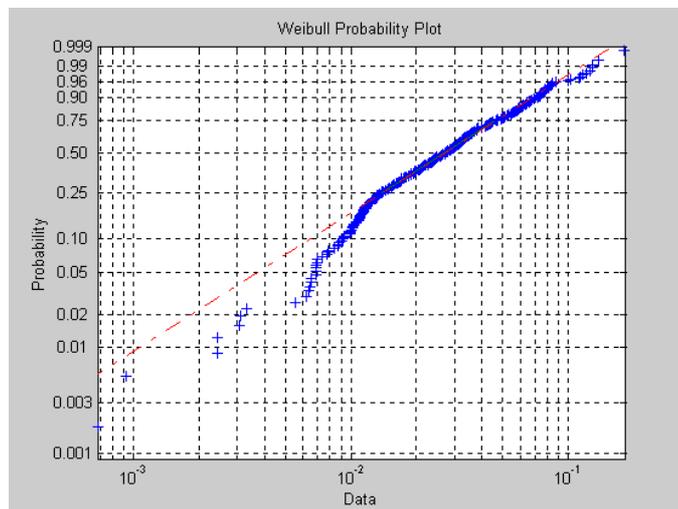

Figure 2. *Hazard plots for pre-WWII and post-WWII expansions and pre-WWII and post-WWII contractions.*

a) *Expansions*

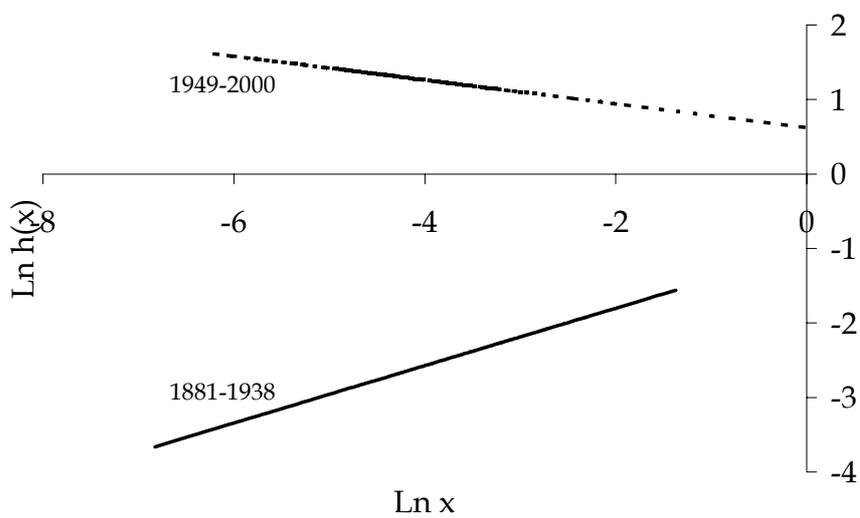

b) *Contractions*

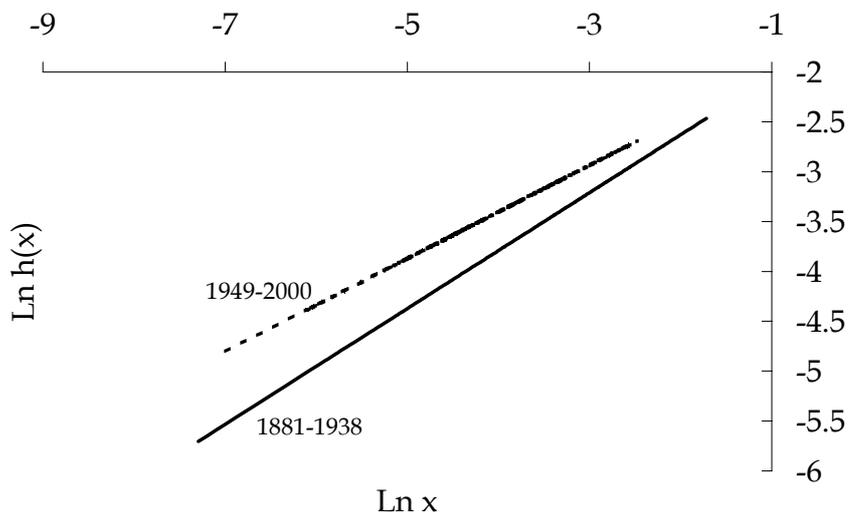